\global\def\draftcontrol{0}

   \def\versionno{2-d Black Hole Thermodynamics}

\catcode`\@=11

\expandafter\ifx\csname draftcontrol\endcsname\relax\global\def\draftcontrol{0}
\fi

{\count255=\time\divide\count255 by 60
\xdef\hourmin{\number\count255}
\multiply\count255 by-60\advance\count255 by\time
\xdef\hourmin{\hourmin:\ifnum\count255<10 0\fi\the\count255}}
\def\draftdate{\number\month/\number\day/\number\year\ \ \ \hourmin }

\newcommand\makepapertitle{\par
  \begingroup
    \renewcommand\thefootnote{\@fnsymbol\c@footnote}%
    \def\@makefnmark{\rlap{\@textsuperscript{\normalfont\@thefnmark}}}%
    \long\def\@makefntext##1{\parindent 1em\noindent
            \hb@xt@1.8em{%
                \hss\@textsuperscript{\normalfont\@thefnmark}}##1}%
     \newpage
     \global\@topnum\z@   
     \@makepapertitle
     \thispagestyle{empty}\@thanks
  \endgroup
  \setcounter{footnote}{0}%
  \global\let\thanks\relax
  \global\let\makepapertitle\relax
  \global\let\@makepapertitle\relax
  \global\let\@thanks\@empty
  \global\let\@author\@empty
  \global\let\@date\@empty
  \global\let\@title\@empty
  \global\let\title\relax
  \global\let\author\relax
  \global\let\date\relax
  \global\let\and\relax
  \def\version{\let\version\@version\@gobble}
}
\def\@makepapertitle{%
  \newpage
   \ifnum\draftcontrol=1 {}
   \version\versionno
   \vskip 3em%
   \else
   \hfill\hbox to 3cm {\parbox{4cm}{\@pubnum}\hss}%
   \vskip 3em%
   \fi
   \begin{center}%
   \let \footnote \thanks
     {\LARGE {\@title}}%
     \vskip 1.5em%
     {\normalsize
       \lineskip .5em%
       \begin{tabular}[t]{c}%
         \@author
       \end{tabular}\par}%
     \vskip 1.5em%
     {\@bstract}%
     \end{center}%
     \vskip 1.5em 
     \@date%
   \par
}

\gdef\@pubnum{}
\def\pubnum#1{%
  \gdef\@pubnum{#1}}

\gdef\@bstract{}
\def\Abstract#1{%
  \gdef\@bstract{%
   \parbox{\textwidth-0pc}{%
   \centerline{\bf Abstract}\penalty1000%
\noindent
\renewcommand\baselinestretch{1.0}%
{#1}}}
}

\def\ps@paper{\let\@mkboth\@gobbletwo%
     \ifnum\draftcontrol=1
        \def\@oddfoot{\hbox to \textwidth{\tiny \versionno \hfil\tiny\draftdate}%
        \hskip -\textwidth \hbox to \textwidth{\hfil\rm\thepage\hfil}}%
     \else\def\@oddfoot{\hbox to \textwidth{\hfil\rm\thepage\hfil}}
     \fi
     \let\@evenfoot\@oddfoot
}

\def\@version#1{\ifnum\draftcontrol=1
\typeout{}\typeout{#1}\typeout{}
\vskip3mm\centerline{\hbox{\fbox{\normalsize{\tt DRAFT -- #1 -- }
                   {\draftdate}}}}\vskip3mm
\fi}
\let\version\@version
\long\def\eqlabel#1{\ifnum\draftcontrol=1
                    \tag@false  
                    \tag*{(\theequation) \hbox to -0.2cm{\hspace{0cm}\small{#1}\hss}}
                    \refstepcounter{equation} 
                    \edef\@currentlabel{\theequation}
                    \ltx@label{#1}          
                    \else
                    \label{#1}
                    \fi
                    }
\let\st@bibitem\@bibitem
\let\st@lbibitem\@lbibitem
\ifnum\draftcontrol=1
  \def\@bibitem#1{%
    \st@bibitem{#1}\a@@label{#1}\ignorespaces}
  \def\@lbibitem[#1]#2{%
    \st@lbibitem[#1]{#2}\a@@label{#2}\ignorespaces}
  \def\a@@label#1{%
    \gdef\a@lab{\smash{\normalfont\small#1}}
    \ifvmode
      \if@inlabel
        \global\setbox\@labels\hbox{%
          \llap{\a@lab\let\a@lab\relax
                \kern\@totalleftmargin\kern\marginparsep}%
          \box\@labels}%
      \fi
    \fi}
\fi

\documentclass[12pt,letterpaper]{article}

\usepackage{amsmath,amssymb,array,calc,rotating,epsfig,psfrag}
\usepackage[nosort]{cite}
\usepackage{color}
\ifnum\draftcontrol=1
\fi

\renewcommand\baselinestretch{1.25}
\renewcommand\section{\@startsection {section}{1}{\z@}%
                                   {-3.5ex \@plus -1ex \@minus -.2ex}%
                                   {2.3ex \@plus.2ex}%
                                   {\normalfont\large\bfseries}}
\renewcommand\subsection{\@startsection{subsection}{2}{\z@}%
                                   {-3.25ex\@plus -1ex \@minus -.2ex}%
                                   {1.5ex \@plus .2ex}%
                                   {\normalfont\normalsize\bfseries}}
\renewcommand\subsubsection{\@startsection{subsubsection}{3}{\z@}%
                                   {-3.25ex\@plus -1ex \@minus -.2ex}%
                                   {1.5ex \@plus .2ex}%
                                   {\normalfont\normalsize\it}}
\renewcommand\paragraph{\@startsection{paragraph}{4}{\z@}%
                                   {-3.25ex\@plus -1ex \@minus -.2ex}%
                                   {1.5ex \@plus .2ex}%
                                   {\normalfont\normalsize\bf}}

\def\revise#1       {\raisebox{-0em}{\rule{3pt}{1em}}%
                     \marginpar{\raisebox{.5em}{\vrule width3pt\
                     \vrule width0pt height 0pt depth0.5em
                     \hbox to 0cm{\hspace{0cm}{%
                     \parbox[t]{4em}{\raggedright\footnotesize{#1}}}\hss}}}}

\def\calm         {{\cal M}}

\def\calo         {{\cal O}}

\def\del          {\partial}

\def\tr           {\mathop{\rm Tr}}

\def\half{{\frac12}}

\def\sqr#1#2{{\vcenter{\vbox{\hrule height.#2pt  
 \hbox{\vrule width.#2pt height#1pt \kern#1pt
 \vrule width.#2pt}\hrule height.#2pt}}}}

\def\a{\alpha}
\def\b{\beta}
\def\r{\rho}

\def\m{\mu}
\def\g{\gamma}
\def\l{\lambda}
\def\n{\nu}
\def\P{\Phi}

\catcode`\@=12

\begin{document}


\topmargin=-0.50in

\newcommand{\be}{\begin{equation}}
\newcommand{\ee}{\end{equation}}
\newcommand{\beq}{\begin{equation}}
\newcommand{\eeq}{\end{equation}}
\newcommand{\ba}{\begin{eqnarray}}
\newcommand{\ea}{\end{eqnarray}}
\newcommand{\nn}{\nonumber}

\def\vol{\bf vol}
\def\Vol{\bf Vol}
\def\del{{\partial}}
\def\vev#1{\left\langle #1 \right\rangle}
\def\cn{{\cal N}}
\def\co{{\cal O}}
\def\IC{{\mathbb C}}
\def\IR{{\mathbb R}}
\def\IZ{{\mathbb Z}}
\def\RP{{\bf RP}}
\def\CP{{\bf CP}}
\def\Poincare{{Poincar\'e }}
\def\tr{{\rm tr}}
\def\tp{{\tilde \Phi}}
\def\Y{{\bf Y}}
\def\te{\theta}
\def\bX{\bf{X}}

\def\TL{\hfil$\displaystyle{##}$}
\def\TR{$\displaystyle{{}##}$\hfil}
\def\TC{\hfil$\displaystyle{##}$\hfil}
\def\TT{\hbox{##}}
\def\HLINE{\noalign{\vskip1\jot}\hline\noalign{\vskip1\jot}} 
\def\seqalign#1#2{\vcenter{\openup1\jot
  \halign{\strut #1\cr #2 \cr}}}
\def\lbldef#1#2{\expandafter\gdef\csname #1\endcsname {#2}}
\def\eqn#1#2{\lbldef{#1}{(\ref{#1})}%
\begin{equation} #2 \label{#1} \end{equation}}
\def\eqalign#1{\vcenter{\openup1\jot
    \halign{\strut\span\TL & \span\TR\cr #1 \cr
   }}}
\def\eno#1{(\ref{#1})}
\def\href#1#2{#2}
\def\half{{1 \over 2}}

\def\ads{{\it AdS}}
\def\adsp{{\it AdS}$_{p+2}$}
\def\cft{{\it CFT}}

\newcommand{\ber}{\begin{eqnarray}}
\newcommand{\eer}{\end{eqnarray}}

\newcommand{\bea}{\begin{eqnarray}}
\newcommand{\eea}{\end{eqnarray}}

\newcommand{\beqar}{\begin{eqnarray}}
\newcommand{\cN}{{\cal N}}
\newcommand{\cO}{{\cal O}}
\newcommand{\cA}{{\cal A}}
\newcommand{\cT}{{\cal T}}
\newcommand{\cF}{{\cal F}}
\newcommand{\cC}{{\cal C}}
\newcommand{\cR}{{\cal R}}
\newcommand{\cW}{{\cal W}}
\newcommand{\eeqar}{\end{eqnarray}}
\newcommand{\lm}{\lambda}\newcommand{\Lm}{\Lambda}
\newcommand{\eps}{\epsilon}


\newcommand{\nonu}{\nonumber}
\newcommand{\oh}{\displaystyle{\frac{1}{2}}}
\newcommand{\dsl}
  {\kern.06em\hbox{\raise.15ex\hbox{$/$}\kern-.56em\hbox{$\partial$}}}
\newcommand{\as}{\not\!\! A}
\newcommand{\ps}{\not\! p}
\newcommand{\ks}{\not\! k}
\newcommand{\D}{{\cal{D}}}
\newcommand{\dv}{d^2x}
\newcommand{\Z}{{\cal Z}}
\newcommand{\N}{{\cal N}}
\newcommand{\Dsl}{\not\!\! D}
\newcommand{\Bsl}{\not\!\! B}
\newcommand{\Psl}{\not\!\! P}
\newcommand{\eeqarr}{\end{eqnarray}}
\newcommand{\ZZ}{{\rm \kern 0.275em Z \kern -0.92em Z}\;}

\def\s{\sigma}
\def\a{\alpha}
\def\b{\beta}
\def\r{\rho}
\def\d{\delta}
\def\g{\gamma}
\def\G{\Gamma}
\def\ep{\epsilon}
\makeatletter \@addtoreset{equation}{section} \makeatother
\renewcommand{\theequation}{\thesection.\arabic{equation}}

\def\be{\begin{equation}}
\def\ee{\end{equation}}
\def\bea{\begin{eqnarray}}
\def\eea{\end{eqnarray}}
\def\m{\mu}
\def\n{\nu}
\def\g{\gamma}
\def\p{\phi}
\def\L{\Lambda}

\begin{titlepage}

\version\versionno

\leftline{\tt hep-th/0402152}

\vskip -.8cm

\rightline{\small{\tt MCTP-04-07}}
\rightline{\small{\tt PUPT-2110}}

\vskip 1.7 cm

\centerline{\bf \Large On Black Hole Thermodynamics of 2-D Type 0A}
\vskip .2cm
\vskip 1cm
{\large }
\vskip 1cm

\centerline{\large Joshua L. Davis${}^1$, Leopoldo A. Pando Zayas${}^1$ 
and Diana Vaman${}^{2}$ }

\vskip .4cm 
\centerline{\it ${}^1$ Michigan Center for Theoretical
Physics}
\centerline{ \it Randall Laboratory of Physics, The University of
Michigan}
\centerline{\it Ann Arbor, MI 48109-1120}
\centerline{\tt joshuald, lpandoz@umich.edu}

\vskip .4cm \centerline{\it ${}^2$ Department of Physics}   
\centerline{ \it Princeton University}   
\centerline{\it Princeton, NJ 08544}  
\centerline{\tt dvaman@feynman.princeton.edu}

\vspace{1cm}

\begin{abstract}

We present a detailed analysis of the thermodynamics  of two dimensional
black hole solutions to type 0A with $q$ units of electric and
magnetic flux. We compute the free energy and derived quantities such as 
entropy and mass for an arbitrary non-extremal black hole. 
The free energy is non-vanishing, in contrast to the case 
of dilatonic 2-d black holes without electric and magnetic fluxes.
The entropy of the extremal black holes is obtained, and we find it to be 
proportional to $q^2$, the square of the RR flux. We compare these
thermodynamics quantities with those from  candidate matrix model duals. 
\end{abstract}



\end{titlepage}




\section{Introduction}\label{intro}

Black hole thermodynamics provides a bridge between the classical and
quantum aspects of gravitational physics. String theory has achieved
moderate success in describing the statistical origin of the 
thermodynamics of some black holes \cite{stbh}. 
The recent proposal for a non-perturbative matrix model description of
two-dimensional type 0 string theory \cite{0att,0ad} opens the
possibility of providing a statistical description of some of the
black holes that appear as solutions to the low energy effective
action \cite{bo,bgv}. 

The thermodynamics of 2-d black holes is very different from its 
higher dimensional counterparts. Notions like the area of the horizon
are simply lacking. There has been, however, extensive work on 2-d
black hole thermodynamics and by now this area is well established
\cite{ew,gp,chiara,mann,mny,lva,frolov,gklm} (see \cite{gr} for a
comprehensive review).

Our aim in this paper is to compute the thermodynamics of the 
2-dimensional black hole of 0A string theory with $q$ units  of
electric and magnetic fluxes. Our analysis yields
a robust expression for the entropy of this class of black holes which
we proceed to compare with the corresponding results provided by matrix
models.  Recently \cite{gtt}, the mass of the extremal black hole in 2-d
type 0A with $q$ units of electric and magnetic RR fluxes was shown to coincide 
with the energy of the deformed matrix model proposed by
Jevicki and Yoneya \cite{jy}. Some other quantities  
were successfully matched in \cite{ulf}. Our analysis points, however,
to a qualitative connection to another matrix model proposed by
Kazakov, Kostov and Kutasov (KKK) \cite{3k}. Some recent work
concerning the KKK model appear in \cite{yin,yi}

The paper is organized as follows. In section \ref{solution} we review the solution
under investigation. Section \ref{adm} is devoted to the calculation of
the ADM mass. Section \ref{thermo} contains our main results which are
explicit expressions for the free energy, entropy and thermodynamical
mass. We also discuss various limits as a way to gain intuition into the
results. In section \ref{matrix}  we compare our results with those
provided by matrix models. In section \ref{conclusion} we summarize
our results and discuss future directions. We
have also included appendix \ref{onshellaction} where we derive the
onshell action and appendix \ref{marginal} where we discuss the
possibility of a marginal deformation corresponding to turning on the
tachyon field and therefore moving into the $\mu\ne 0$ space in the
matrix model side.

\section{The black hole solution}\label{solution}

The low energy effective action for 2-d type 0A string theory in the
presence of RR flux is \cite{0ad}:
\be
\eqlabel{ea}
\begin{split}
S=\int d^2 x\sqrt{-g}\bigg[\frac{e^{-2\Phi}}{2\kappa^2}&\left(
\frac{8}{\a'}+ R + 4(\nabla\Phi)^2 -f_1(T)(\nabla T)^2
+f_2(T)+\ldots\right)  \\
&-\frac{2\pi\a'}{4}f_3(T)(F^{(+)})^2 -\frac{2\pi\a'}{4}f_3(-T)(F^{(-)})^2
+\ldots\bigg],
\end{split}
\ee
where $f_i(T)$ are functions of the tachyon field $T$. It is convenient to
dualize the RR field strengths  following \cite{thompson}. Moreover,
in the sector with equal number $q$ of electric and magnetic D0 branes the
action reduces to \cite{gtt}:
\be
\eqlabel{leea}
S=\int d^2 x \sqrt{-g}\bigg[e^{-2\Phi}\left(c+R+ 4 (\nabla\Phi)^2 -
(\nabla T)^2 + \frac2{\a'}T^2\right)+ \L(1+2T^2)\ldots\bigg], 
\ee
where $c=8/\a'$ and $\L=-q^2/(2\pi\a')$ and we work in units where 
$2\kappa^2=1$.

A particularly simple class of solution to the equations of motion
corresponds to $T=0$. In this case the action becomes a 2-d dilaton
gravity with nontrivial cosmological constant. Black hole solutions to
such action have been presented in \cite{bo,bgv,gtt}\footnote{We have
rescaled the solution with respect to the standard presentation in the
literature \cite{bgv,gtt}: $t\to
t\sqrt{\frac{4}{c}}$ and $\phi \to \phi \sqrt{\frac{c}{4}}$. This
rescaling guarantees that the metric asymptotes to the flat metric.} . 
\be
\eqlabel{sol}
ds^2=l(\phi)\,dt^2 + \frac{d\phi^2}{l(\phi)},
\ee
where
\be
l(\phi)=1-\frac{4}{c}e^{\sqrt{c}\phi}\left(\frac14 \Lambda \, \sqrt{c}\, \phi +m \right),
\ee     
and the dilaton $\Phi=\sqrt{c}\,\p/2$. For $\L<0$ the generic solution 
looks like a
2-d version of the Reissner-Nordstrom black hole, that is, these are
charged solutions with two horizons \cite{bgv}.  

In the region $\phi\to -\infty$ this solution asymptotes to the
linear dilaton solution:
\be
\eqlabel{lineardil}
ds^2= -dt^2 +d\phi^2, \qquad \Phi=\sqrt{\frac{c}{4}}\;\p.
\ee
The near horizon geometry generically looks like 2-d Rindler
space with metric $-x^2 dt^2 +dx^2$. This can be seen by expanding
$l(\phi)$ to first order near the
outer horizon and then introducing $x\sim \phi^{1/2}$. For extremal
black holes the linear term vanishes yielding $l(\phi) \sim \phi^2$
which means that the near horizon geometry is $AdS_2$ with metric
$-\phi^2 dt^2 +d\phi^2/\phi^2$ \cite{bgv}. The fact that the extremal black hole
interpolates between ``flat space'' and $AdS_2$ was noted in \cite{andy}
where this analogy with the D3  brane background was pushed to a
proposal for $AdS_2/CFT_1$.

Even though the solution is obtained with vanishing tachyon and
therefore zero Liouville potential we argue in appendix \ref{marginal}
that there is a marginal deformation in the direction of nonzero
$\mu$. The existence of such deformation encourages us to believe that
there is a well-defined description of these black holes in terms of a
matrix models.

\section{The ADM mass of a 0A 2-d black hole}\label{adm}

The question of mass in 2-d dilaton gravity has been answered in a very
general context. In this section we follow an account due to Mann
\cite{mann} which generalizes previous work of Frolov \cite{frolov} 
(see \cite{lva} for an alternative approach). 

The starting point is an action of the general form
\be
\eqlabel{mannaction}
S[g,\P]=\int d^2 x \sqrt{-g}\bigg[D(\P)\,R
+H(\P) g^{\m\n}\del_\m\P \del_\n \P  +V(\P,\P_M)\bigg],
\ee
where $\P_M$ denotes other types of matter. In the analysis of
\cite{mann} $V(\P,\P_M)$ is
restricted to have no metric dependence. The presence of a kinetic term
for the tachyon in (\ref{leea}) would naturally prevent us from
simply borrowing the results of \cite{mann}. However, for configurations of
constant tachyon the potential is indeed independent of the metric and
the result of \cite{mann} applies. Note that for a constant tachyon the
action (\ref{leea}) essentially becomes:
\be
\eqlabel{action1} 
S=\int d^2 x \sqrt{-g}\bigg[e^{-2\P}\left(R+ 4(\nabla\P)^2 +c\right)
+ \Lambda\bigg]
\ee
and the functions $D(\P), H(\P), V(\P)$ from the generic action 
(\ref{mannaction}) can be identified as
\be
D(\P)=e^{-2\P},\; H(\P)=4e^{-2\P},\;
V=\Lambda+ c\;e^{-2\P}.
\ee
A generic action such as (\ref{mannaction}) admits a topologically conserved
current \cite{mann, mny}, 
\be
\eqlabel{cur}
S_\mu= T_{\mu\nu}\epsilon^{\nu\rho}\partial_\rho F
\ee
where $T_{\mu\nu}$ is the corresponding stress-energy tensor, provided that
\be
F=F_0\int\limits^{\P} ds
\;D'\,\exp\left(-\int\limits^s dt\,\frac{H(t)}{D'(t)}\right).
\ee
In particular, for us $F$ is proportional to the dilaton $\P$.
The proportionality constant $F_0$ is fixed 
from the condition that for large $x$
\be
\eqlabel{norma}
\lim \frac{d F}{ dx} \longrightarrow 1.
\ee 
The  current (\ref{cur}) can be used to define a mass
$\calm$ independently of the
existence of a time-like Killing vector via\footnote{The expression for
  $\calm$ in \cite{mann} differs from equation (\ref{mass})by a factor of two. We have fixed this
  overall coefficient using the ADM mass for the 2-d black hole
  discussed by Witten in \cite{ew}. Our normalization also agrees with
  the ADM mass of \cite{gp} in the case of vanishing RR flux.}
$S^\mu=\epsilon^{\mu\nu}
\partial_\nu \calm$
\be
\calm=F_0 \bigg[\int\limits^\P ds D'(s)\, V(s)\,
\exp\left(-\int\limits^s dt\,\frac{H(t)}{D'(t)}\right) -(\nabla D)^2 
\exp\left(-\int\limits^\P d t\, \frac{H(t)}{D'(t)}\right)\bigg].
\ee
For the solution at hand we obtain that:  
\be
\calm=4 F_0e^{-2\P}\bigg[(\nabla\P)^2 -\frac{c}4+\frac12\Lambda \,\,
\Phi e^{2\P}\bigg]
\ee
and $F_0=1/\sqrt{c}$. Evaluated on the solution (\ref{sol}) 
we find that the mass is 
\be
\eqlabel{mass}
\calm=\frac{4}{\sqrt{c}}\;m.
\ee
A similar expression for the mass was obtained in \cite{bgv,gtt} and 
justifies the notation for the constant $m$ in the
general solution. We disagree however with the ADM mass expression 
reported  in \cite{gtt}: $\frac 2{\sqrt c} m + \frac{\Lambda}{2\sqrt
c}$. The second term is absent from our evaluation of the ADM 
mass. We will shortly re-derive the same black hole mass via a thermodynamical
analysis.

\section{Black hole thermodynamics}\label{thermo}

This section is dedicated to studying the thermodynamics
of the solution presented in section \ref{solution}. We
will first discuss the temperature and dilaton charge associated with
this class of 2-d black holes before moving on to the free
energy and derived quantities. 

We will be able to recover from our results the 
thermodynamics of dilatonic 2-d black holes with vanishing cosmological
constant \cite{gp}. We have exact results for an arbitrary non-extremal
black hole, but for the sake of developing some intuition into the
behavior of these black holes we address the near-extremal case 
separately. Finally, being concerned with a possible matrix model description
along the lines of \cite{gtt}, we explore the large RR flux  limit,
which is realized as the near extremal limit on the gravity side.

\subsection{Temperature}

For a metric of the form (\ref{sol}), the corresponding temperature can
be computed from the condition that the Euclidean counterpart does not 
have conical
singularities near the outer horizon\footnote{This could be easily
seen by introducing $r=l^{1/2}$ such that $dr=\frac12 l^{-1/2}l' d\p$
and 
\be
ds^2 = \frac{4}{l'^2}\left(dr^2+\frac14 l'^2 r^2 dt^2\right).
\ee}:
\be
\eqlabel{t00}
T=\frac{1}{4\pi} |l'(\p)|_{\p=\p_{h}}.
\ee
We need to evaluate the above expression at the largest root of
$l(\p_h)=0$. The location of the horizon is dictated by the 
equation:
\be
1-\frac{4}{c}e^{\sqrt{c}\p_h}\left(\frac14\, \L\sqrt{c}\,\p_h +m\right)=0. 
\eqlabel{horizon}
\ee
The solution to this equation is given via the Lambert function which
by definition satisfies $W(z)\exp(W(z))=z$: 
\be
\eqlabel{phi}
\p_{h}=-\frac{4m}{\L\,\sqrt{c}}+\frac1{\sqrt{c}} W\left(\frac{c}{\L}\, e^{4m/\L}\right).
\ee
Equations (\ref{t00}) and (\ref{phi}) are sufficient to
evaluate the temperature of a general non-extremal black hole:
\be
\eqlabel{t}
T=\frac{\sqrt{c}}{4\pi} |1+\frac{\L}{c}\exp\left(-4\frac{m}{\L}+W(\frac{c}{\L}e^{4m/\L})\right)|.
\ee
To gain intuition into the expression for the temperature, let us
consider the extremal and near extremal case. 

The extremal black
hole has zero temperature. The position of the horizon and its mass are:
\be
\eqlabel{extremal}
\p_0=-\frac1{\sqrt{c}}-4\, \frac{m_0}{\L\sqrt{c}}=\frac{1}{\sqrt{c}}
\ln \left(-\frac{c}{\L}\right), \qquad 
m_0=-\frac14\,\L \bigg[1+\ln\left(-\frac{c}{\L}\right)\bigg].
\ee
The above equations (\ref{extremal}) imply the mass of the extremal black hole can become 
negative for large enough RR flux. In two dimensions there is an
analogue of Witten's proof of the positivity of the ADM mass
due to Park and Strominger \cite{positive}. Thus, in view of the form
for the ADM mass (\ref{mass}) and the expression for the extremal
black hole (\ref{extremal}) we conclude that amount of flux has an
upper bound: 

\be
\eqlabel{bound}
q^2<16\pi e.
\ee
For a black hole with ADM mass slightly higher than the extremal mass
which correspond to parameters $m$ in the range 
$m=m_0+\d m$ with $|\d m/m_0| \ll 1$, the position of the outer horizon becomes:
\be
\p_{h}=\frac1{\sqrt{c}}\ln
\left(-\frac{c}{\L}\right)+\frac{2\sqrt{2}}{\sqrt{c}}\left(-\frac{\d m}{\L}\right)^{1/2}
-\frac4{3\sqrt{c}}\frac{\d m}{\L} + \calo\left(\left(-\frac{\d
      m}{\L}\right)^{3/2}\right).
\ee
Thus, the horizon is pushed outward by adding a small amount of
matter. This behavior is as in \cite{ew}.
Similarly, the temperature  corresponding to this 
near-extremal black hole is: 
\be
\eqlabel{temperature}
T=\frac{\sqrt{c}}{\sqrt{2}\pi}\left(-\frac{\d m}{\L}\right)^{1/2}.
\ee

\subsection{Dilaton Charge}

In two dimensions there is a remarkable ambiguity in the choice of the
dilaton charge. Any current of the form, $j_\mu = -\epsilon_\mu^\nu
\nabla_\nu f (\P)$, is conserved by symmetry arguments without the
involvement of the equations
of motion and is therefore  topological. Namely,  
\be
\nabla_\mu j^\mu = -\epsilon^{\mu\nu} \left[ f^{\prime\prime}
(\P) \nabla_\mu \P \nabla_\nu \P + f^\prime (\P) \nabla_\mu \nabla_\nu \P \right].
\ee
The right hand side of the previous equation 
is clearly vanishing, irrespective of the choice of the function $f(\Phi)$.

The topological charge associated with the above current is the flux 
of this current
through a space-like slice, $\Sigma$, which stretches from the horizon
to some cut-off wall:
\bea
D & = &\int_\Sigma d\Sigma n^\mu j_\mu \nonumber \\
  & = & -\int_{\phi_W}^{\phi_0} d\phi \,  \sqrt{g_{\phi\phi}} n^t
  g_{tt} 
\epsilon^{t\nu} \nabla_\nu f(\Phi)\nonumber \\
  & = & -\int_{\phi_W}^{\phi_0} d\phi \, f^\prime (\P).
\eea
We take the canonical choice for $f(\Phi)$ such that the dilaton charge
is $D=e^{-2\Phi}$, that is, the function multiplying
the Ricci scalar in the action. This choice 
facilitates the comparison of our results with other relevant calculations in
the literature.

\subsection{Thermodynamic relations}

 \subsubsection{The free energy}

In this section we evaluate the on-shell Euclidean action corresponding to 
the black
hole solution (\ref{sol}). The action has been derived in appendix
\ref{onshellaction} and is given by:
\be
\eqlabel{action}
I_{onshell} = \int_{\cal M} \sqrt{g} \Lambda 
+2  \int_{\partial {\cal M}} \sqrt{h} e^{-2\Phi} 
\left( K -2 n^a \nabla_a \Phi \right). \\
\ee
Our overall strategy is to  extract the thermodynamic quantities associated 
with this
solution following the general approach of \cite{gh}. 
We will use a
concrete analysis due to  \cite{gp}\footnote{
Another paper which attempts to provide a rather general framework
for evaluating the thermodynamical mass and the entropy of a dilaton-gravity
solution is \cite{chiara}. The point of view embraced by the authors of
\cite{chiara} is that the on-shell dilaton-gravity 
action associated with a static solution can always be expressed as
$$I_{onshell}=-\int  dt \int_{horizon}^{spatial\;infinity}
dx\partial_x [4 e^{-2\Phi} l(\Phi)\partial_x \Phi+
e^{-2\Phi}\partial_x l(\Phi)].$$ Furthermore, the second term when evaluated
at the horizon yields the entropy, and when evaluated at infinity gives
the thermodynamical mass. This observation is based on the fact that
$I_{onshell}=\beta F=\beta E -S$. The role played by the first term of the
onshell action is to account for the chemical potential associated with the 
dilaton charge. 
A difference between this approach and the one we used is that for us the
Einstein-Hilbert action is supplemented with a boundary term that ensures
that the variational principle is satisfied. Therefore the boundary
terms in (\ref{action}) are evaluated on the boundary of the 2-d manifold
that is the Euclidean black hole solution, which is to say that the {\it only} 
contribution coming from the boundary terms arises from spatial
infinity.} and derive the on-shell action 
in terms of measurable/observable quantities. The basic setup is that of making
physical observations at the wall of a box that serves as the boundary
of space-time. At the wall we can measure the value of the dilaton charge
$D_W$ and the temperature $T_W$ there. In terms of these observable
variables the free energy, entropy, energy and 
dilaton chemical potential can be obtained from the on-shell action $I$ as:
\be
\eqlabel{thermalq}
F=T_W I,\qquad S=-\frac{\partial F}{\partial T_W}, \qquad E= F+T_W S,
\qquad \psi=-\frac{\partial F}{\partial D_W}.
\ee
For a metric of the form (\ref{sol}) the quantities related to the
curvature are
\be
R=-\del^2_{\p}l, \qquad \G^\p_{tt}=-\frac{l}2\del_\p l,\qquad
\G^t_{t\p}=-\G^\p_{\p\p}=\frac{1}{2l}\del_\p l.
\ee
Considering a foliation given by a unit vector in the $\p$ direction
$n^\p=\sqrt{l}$, the extrinsic curvature of the surface (in this case, curve) 
orthogonal to the foliation is: 
\be
K_{tt}=h_{tt}\G^t_{t\p}n^\p=\frac{\sqrt{l}}{2}\del_\p l, 
\qquad K=\G^t_{t\p}n^\p=\frac{1}{2\sqrt{l}}\del_\p l.
\ee
With these ingredients the onshell  action becomes:
\bea
I &=& \b\,\L(\p_W-\p_h)+\beta e^{-\sqrt{c}\,\phi_W} (-2l(\phi_W)\,\sqrt{c}+l'(\phi_W) )\nonumber\\
&=&  \b\,\L(\p_W-\p_h)-\beta\,\sqrt{c}\, e^{-\sqrt{c}\,\phi_W}
\left(l(\phi_W)+1+\frac{\Lambda}{c}\,e^{\sqrt{c}\,\p_W} \right),
\label{onshell} 
\eea
where $\beta$ is the inverse temperature of the 2-d black hole.
Using the Tolman relation  which relates the temperature at 
the wall $T_W$
to the black hole temperature $T$ 
\be
\eqlabel{tolman}
T_W=T\frac{1}{\sqrt{l(\phi_W)}},
\ee
where 
\be
\eqlabel{t0}
T=\frac{\sqrt{c}}{4\pi}\bigg|1+\frac{\Lambda}{c}D_h^{-1}\bigg|,
\ee
and by expressing the parameter $m$ of the 2-d black hole solution 
as a function of the position of the horizon $m(\p_{h})$
\be
\eqlabel{m}
m(\p_h)=\frac{c}{4}\; D_h+\frac{\Lambda}4\;\ln D_h,
\ee
we arrive at the following expression of 
the on-shell action (\ref{onshell}):
\be
I_W= \frac1{T}\,\L(\p_W-\p_h)-\frac{D_W}{T}\left(1+ \frac{T^2}{T_W^2}
+\frac{\L}{c\,D_W}\right).
\ee
This expression of the free energy
is not yet ready to evaluate the thermodynamic quantities
according to (\ref{thermalq}) because it is not expressed exclusively in terms
of observable quantities $(T_W,D_W)$ and parameters of the theory
$(c,\L)$. In particular, we would like to substitute the 
dilaton charge at the horizon  $D_h=e^{-2\Phi_h}$ 
by an alternative expression dependent on the observables $(T_W,D_W)$ and
$(c,\L)$. 
This may be achieved by using the Tolman relation: 
from (\ref{tolman}) and (\ref{t0}) we find an expression containing $D_h$ as:
\be
\eqlabel{tw}
T_W=\frac{\sqrt{c}}{4\pi}\frac{1+\frac{\L}{c}D_h^{-1}}
{\sqrt{1-\frac{D_h}{D_W}-\frac{\L}{c\,D_W}\ln
\frac{D_h}{D_W} }}.
\ee
This equation should be viewed as an equation for the implicit dependence
of the dilaton charge at the horizon on the temperature at the wall.

We have thus determined the free energy of the 
2-d black hole: 
\be
F=-\frac{\L}{\sqrt{c}} \,\frac{T_W}{T}\ln\frac{D_W}{D_h}-
\sqrt{c}\,D_W\frac{T_W}{T} 
\left(1+\frac{T^2}{T_W^2}+\frac{\Lambda}{c\,D_W}\right),
\ee
where $T$ and $D_h$ should be understood as functions of $(T_W,D_W)$
following from (\ref{t0}) and (\ref{tw}).

 \subsubsection{Thermodynamics at zero RR flux}

Let us consider the case of zero cosmological constant, that is,
the solution in the absence of RR flux. The solution for the dilaton at
the horizon in terms of physical quantities that follows from (\ref{tw})
is:
\be
D_h=D_W\,\left(1-\frac{c}{16\pi^2\,T_W^2}\right).
\ee
This value of the dilaton allows us to identify $m(\phi_h)$ as:
\be
m(D_W,T_W)=\frac{c\,D_W}{4}\left(1- \frac{c}{16\pi^2\,T_W^2}\right).
\ee
With these ingredients we find the free energy, entropy and energy
of the zero RR flux 2-d black hole:
\bea
F&=&-4\pi\, D_W\left(T_W+\frac{c}{16\pi^2\, T_W}\right), \nonumber \\
S&=&4\pi \,D_W \left(1-\frac{c}{16\pi^2\,T_W^2}\right),\nonumber \\
E&=&-8\pi \,D_W\frac{c}{16\pi^2\, T_W}.
\eea
These quantities coincide precisely with the ones obtained in
\cite{gp} upon the identification of the temperature at the horizon,
$T_c$ in the notation of \cite{gp}, with  $\sqrt{c}/4\pi$.
The flat space linear dilaton subtraction regularizes the divergent 
quantities. In particular, one finds that this solution has vanishing free
energy, that the mass  (obtained from the regularized
energy) coincides with the ADM mass $M=E-E_{flat~space}=
8\pi D_W T(1-\frac{T}{T_W})=4\pi T D_h$, and that the
entropy is $S=M/T$.

 \subsubsection{Extremal black hole}
In this section we extract some of the thermodynamic properties of the
extremal black hole. We will approach these quantities by considering
a near extremal black hole. 
In the near extremal limit the leading order solution to (\ref{tw}) has
to take the form of (\ref{extremal}), that is 
\be
\p_h=\frac1{\sqrt{c}}\ln\left(-\frac{c}{\L}\right) + \d\p_h.
\ee
We can solve to first order in the leading correction
\be
\eqlabel{dp}
\d\p_h=-4\pi \frac{T_W}{c}\left(1+\frac{\L}{c\,D_W}\left(1
+\ln(-\frac{cD_W}{\L})\right)\right)^{1/2}.
\ee
This is enough to evaluate the free energy 
\be
F=\frac{T_W}{T}\left(-\frac{\L}{\sqrt c}\ln(-\frac{c\,D_W}{\L})-\sqrt{c}\,D_W
  - \frac{\L}{\sqrt{c}} -\L\d\p_h\right) -\sqrt{c}\,D_W\frac{T}{T_W},
\ee
where $\d\p_h$ is given by (\ref{dp}) and 
\be
\frac{T}{T_W}=\left(1+\frac{\L}{c\,D_W}
\big[1+\ln(-\frac{c\,D_W}{\L})\big]\right)^{1/2}.
\ee
Note that this ratio is independent of the temperature at the wall.
Hence when differentiating the free energy with respect to $T_W$ 
in (\ref{thermalq}) the only term that contributes is the one proportional 
to $\d\p_h$. 
We find that the entropy of the extremal 2-d black hole with RR flux
is simply
\be
\eqlabel{entropy}
S=-\frac{4\pi\,\L}{c}=\frac14 q^2.
\ee
This value of the entropy is natural to identify with the entropy of the
extremal black hole with $q$ units of electric and magnetic RR fluxes. 

The perspective of working with a solution with non-vanishing cosmological
constant can be interchanged with that of discussing a particular solution
with (equal) constant electric and magnetic fluxes.
The advantage which comes from this 
latter point of view is that we can now justify choosing
to regularize the divergent quantities by subtracting the linear dilaton flat
space as a bona fide solution with vanishing fluxes, just as
Gibbons-Hawking treated the Reissner-Nordstrom black hole in the
thermodynamical approach \cite{gh}. 
Let us now consider the thermodynamical energy, which after a suitable
subtraction of the energy of the reference linear dilaton flat space 
background, we would like to
eventually identify with the mass of the black hole. In the large
$D_W$ limit, corresponding to moving the position of the wall to infinity,
 the energy computed from (\ref{thermalq}) is
\be
E=-2\sqrt{c}\,D_W -\frac{\L}{\sqrt{c}} \ln D_W + \frac{\L}{\sqrt{c}}
\left(\ln(-\frac{\L}{c})-1\right).
\ee
It is worth saying that the chosen reference background subtraction will 
remove only the leading divergence
from this expression and we were unable to find another subtraction procedure
which would remove both divergences at the same time. 
We discard the second divergence on the basis that it is an infinite
volume factor. One  remains with a finite part which can be identified with the 
thermodynamical mass of the near-extremal solution
\be
M=\frac{\Lambda}{\sqrt{c}}\left(\ln(-\frac{\L}{c}) -1\right)=
-\frac{q^2}{4\pi \sqrt{2\a'}}\left(\ln\frac{q^2}{16\pi}-1\right).
\ee
This can be seen to coincide with the ADM mass in the extremal case.

Finally, the chemical potential is 
\be
\psi=2\sqrt{c}.
\ee
\subsubsection{Thermodynamics of an arbitrary non-extremal OA 2-d black hole} 

Let us return to the on-shell action of the 0A solution with equal number 
of electric and magnetic fluxes. We saw that the presence of these fluxes 
manifests as a negative cosmological constant term in the action, and it 
is this
term which yields the only bulk contribution to the on-shell action.
To see more distinctly the source of the various terms in the thermodynamical
potentials, let us place a marker, a coefficient $\alpha$ in front of the 
bulk on-shell action. The free energy is then
\be
F = \alpha \frac{\L}{\sqrt{c}} \,\frac{T_W}{T}\ln\frac{D_W}{D_h}-
\sqrt{c}\,D_W\frac{T_W}{T} 
\left(1+\frac{T^2}{T_W^2}+\frac{\Lambda}{c\,D_W}\right),  \qquad
\alpha = -1.
\ee
In the limit where we take the position of the wall to infinity,
we have $D_W\rightarrow \infty, T_W\rightarrow T$, while keeping the
temperature of the black hole $T$ (and thus $D_h$) finite, the free 
energy becomes
\be
F=-2\sqrt{c} D_W -\frac{\alpha \Lambda}{\sqrt {c}}\ln(\frac {D_W}{D_h}) -
\frac{\Lambda}{\sqrt{c}}
+{\cal O}(D_W^{-1}).
\ee
Differentiating (\ref{tw}) with respect to  $T_W$ on both sides one obtains 
an equation
for the derivative of the dilaton charge at the horizon with respect to  
the temperature
at the wall: $d D_h/d T_W$. By substituting it into (\ref{thermalq}),
we obtain the 2-d black hole energy
\bea
E&=&\frac{\sqrt{D_W (c(D_W-D_h)+\Lambda \ln(D_W/D_h))}}{2\Lambda c D_W -
4 c \Lambda D_h +2\Lambda^2 \ln (D_W/D_h) - c^2 D_h^2 -\Lambda^2}
\bigg(-2\Lambda c D_W + 3 \Lambda c D_h \nonumber\\
&+&\Lambda^2 (1+\alpha)\ln(D_h/D_W)
-\alpha \Lambda c D_h - \alpha \Lambda^2\bigg).\nonumber\\
\eea
To find the mass of the non-extremal 2-d black hole, we take as usual
the limit $D_W\rightarrow\infty$ in the expression of the energy, while 
keeping the value of the dilaton charge at the horizon fixed
\be
E=-2\sqrt{c} D_W +\frac{\alpha\Lambda}{\sqrt{c}}\ln(\frac{D_W}{D_h})
-\alpha \sqrt{c}
D_h - (1+\alpha)\frac{\Lambda}{\sqrt{c}} +{\cal O}(D_W^{-1}) \label{neenergy}
\ee
As before, as the wall is pushed to infinity, the energy diverges. The 
leading divergence - the first term in (\ref{neenergy}) -
is canceled by the linear dilaton flat space background subtraction, leaving
us with another divergence, proportional to a volume factor $\Lambda\Phi_W$.
Discarding this term as well, one is left with a 
finite expression which we identify with the thermodynamical mass $M$ of the
2-d black hole. With the numerical coefficient $\alpha=-1$, we find
\be
M=\sqrt{c} D_h + \frac{\Lambda}{\sqrt c} \ln D_h = \frac {4}{\sqrt c} m
\ee
that the thermodynamical mass coincides, as expected, with the ADM mass.
The entropy expressed in terms of the observables ($T_W, D_W$) is 
\be
S=4\pi D_h
\ee
with $D_h$ given by (\ref{tw}). Note that since 
as the wall of the box is 
taken to infinity we keep $D_h$ fixed, the entropy remains constant, 
equal to $4\pi D_h$. 

It might come as a surprise that we get the same answer for the entropy
in terms of the dilaton charge at the horizon for both dilatonic 
2-d black holes with vanishing and non-vanishing cosmological constant.
This result is in fact quite universal for 2-d black holes, as shown by
\cite{gklm}, and our calculation exactly matches the entropy derived
by  \cite{gklm}. It is also worth stressing the robustness of the value
of the entropy. Although it might not be clear from the explicit
calculations, the entropy (as opposed to other thermodynamical
quantities) is insensitive with respect to the position of the wall. 

Also note that the results at zero RR-flux are directly obtained from our
final expressions by taking $\Lambda=0$. It is interesting to notice the
role played by the on-shell bulk (volume term) 
action: with $\Lambda=0$ the on-shell action has 
only boundary terms, which are therefore responsible for the thermodynamical
mass $\sqrt{c} D_h$; computing first the thermodynamical mass at 
$\Lambda\neq 0$ as we did in this subsection and then subsequently setting
$\Lambda=0$ in the final result, one observes that the thermodynamical
mass at vanishing cosmological constant originates entirely in the bulk term
of the on-shell action. 

Let us  summarize our results for the thermodynamics quantities of 
2-d black holes, in terms of the dilaton charge at the horizon $D_h$,
in the following table: 
\begin{table}[hbt] 
\begin{center}     
\begin{tabular}[h]{|c||c|c|c|c|c|}     
\hline    
$ {}$& $M$ &$T$ & $S$ &$F $& $\psi$\\    
\hline      
$\Lambda=0$&$\sqrt{c} D_h$&$\frac{\sqrt{c}}{4\pi}$&$4\pi D_h $& $0 $&$2\sqrt{c}$\\     
\hline\hline     
$\Lambda\neq 0$&$\sqrt{c} D_h +\frac{\Lambda}{\sqrt c}\ln D_h$&$ 
\frac{1}{4\pi \sqrt{c}D_h}
|cD_h+\Lambda|$&$4\pi D_h$&$\frac{\Lambda}{\sqrt c}(\ln D_h-1)$&$2\sqrt{c}$\\
\hline      
\end{tabular}     
\caption{Thermodynamic properties of 2-d black holes in type 0A. \label{table}}
\end{center}     
\end{table} 

\subsubsection{Near extremal thermodynamics}
Given the bound we found for the number of D0 branes (\ref{bound}) we
cannot try to make contact with the matrix model results where the
amount of flux is considered large. Rather a similar limit can be
realized in the black hole by considering the near-extremal limit:
\be
\Delta M/M_0 \equiv \epsilon \ll 1
\ee 
given that $M_0$ is proportional to the flux. 

In this limit we arrive at the following expansions:
\bea
T&=&\frac{\sqrt 2|\Lambda| \sqrt{1+\ln(-c/\Lambda)}}{4\pi\sqrt c}\sqrt\epsilon
+\frac{\Lambda(1+\ln(-c/\Lambda))}{6\pi\sqrt{c}}\epsilon+\dots\\
S&=&4\pi\left(\frac{|\Lambda|}{c}+\frac{\sqrt 2\Lambda\sqrt{1+\ln(-c/\Lambda)}}
{c}\sqrt\epsilon +\dots\right)
\eea
By approximating the temperature with the first term in the expansion, we 
find that the entropy equals the extremal 2d black hole entropy 
plus a correction $\Delta S\propto \Delta M/T$,
even though
we should caution that the temperature is mass-dependent according to
the previous set of equations. Also, the correction to the extremal black 
hole entropy respects the Cardy formula $\Delta S/S\propto \sqrt{\frac{\Delta M}{M_0}
}$.
Similarly, by defining the free energy with respect to the extremal black hole
$\Delta F=\Delta M - T\Delta S$, one finds 
that $\Delta F= -\Delta M/T$, with the temperature of the black 
hole again given by the leading term in (\ref{temperature}). 

\section{Comparison with matrix models}
\label{matrix}

Our main motivation for studying the thermodynamics of 2-d
black holes in type 0A is the possibility of the existence of a
statistical foundation based in matrix models. In this section we
will therefore explore the extent to which a connection can be made
between the black hole as solutions of the low energy supergravity
action and a dual matrix model. 

The natural place to start would be the matrix model for type 0A
discussed in \cite{0ad}. The matrix model in question is described by
a system of $N$ decoupled non-relativistic fermions moving in two
dimensions with angular
momentum related to the RR flux $q$
\be
\eqlabel{potential}
V(\lambda)=-\lambda^2+\frac{q^2-1/4}{\l^2}.
\ee
This is the deformed matrix model of Jevicki and Yoneya \cite{jy} and
has been extensively studied (see for example \cite{jy0,jycom})
\footnote{
As explained in \cite{0ad}, in the presence of a Liouville
potential only one type of brane is physical. That is, the above
potential describes a system where only electric or magnetic branes
are present. The solution discussed in this paper corresponds to both
electric and magnetic charges being turned on, which is possible {\it only}
for a vanishing tachyon background.}.
With the string coupling constant in the deformed matrix model of the order 
$g_s \sim 1/q$, up to one loop (in the $1/q$ expansion) the free energy 
\cite{jycom} is 
\be
{\cal F} = - {1\over 8\pi} q^2 \log {q^2  \over L^4} 
+ {1 \over 48\pi} \left[ 1 + (2\pi T)^2 \right] \log {q^2  \over L^4} \ldots
\ee
where $T$ is the temperature and $L$ is an IR cut-off. The first 
non-vanishing contribution to the entropy is one-loop
\be
S = - {\pi\over 12} T \log {q^2  \over L^4}.
\ee

In a recent paper \cite{gtt}, it was observed that the ADM mass of the 0A
2-d {\it extremal} black hole matches the energy of the ground state of 
the deformed 
matrix model. 
Our thermodynamical analysis reveals that the entropy of the extremal
2-d black hole $S_{extremal}=4\pi|\Lambda|/c= q^2/4$ does not match the 
entropy of the deformed Jevicki-Yoneya matrix model.   

Another matrix model considered as a candidate for describing 2-d black
holes  is the matrix model of
Kazakov, Kostov, and Kutasov (KKK) \cite{3k}.
This matrix model involves summing over all possible $U(N)$ twists
around the Euclidean time circle. The partition function can be written 
in terms of a sum of Gibbs partition functions over $SU(N)$ representations
\be
Z_N(\beta,\lambda)=\sum_r \int [D\Omega]\chi_r (\Omega^\dagger)
\exp(\sum_{n\in {\bf Z}} \lambda_n tr(\Omega^n)) Tr_r e^{-\beta H_r}
\ee
where $\chi_r(\Omega^\dagger)$ is the Weyl character, $H_r$ is the 
Hamiltonian in the representation $r$
\be
H_r=P_r\sum_{k=1}^N \bigg(-\frac 12 \partial_{x_k}^2 - \frac 12 x_k^2\bigg)
+\frac 12 \sum_{i\neq j} \frac{\tau_{ij}^r \tau_{ij}^r}{(x_i-x_j)^2}
\ee
and $x_i$ are the eigenvalues of the matrix with the inverted harmonic
oscillator potential. The matrices $\tau_{ij}^r$ are the $SU(N)$ generators.
The free energy of the KKK matrix model has the form
\be
F=\frac{1}{2\pi R}\left( \frac 14 (2-R)^2 \lambda^{4/(2-R)} -
\frac{R+R^{-1}}{48}\ln(\lambda^{4/(2-R)})+\sum_{h=2}^\infty
f_h(R)\lambda^{4(1-h)/(2-R)}\right) 
\ee
where $R$ is the radius of the compactified time circle, and is therefore 
related to the temperature by $2\pi R=1/T$.
In the critical theory, where the central charge constraint requires $R =
{3\over2}$, the genus zero contribution to the free energy is of the order
$\l^{2 \over (R-2) } \sim M \sim {1\over g^2_s}$.
 
This model has also a large entropy which is assumed to be associated with the
non-singlet sector of the matrix theory \cite{3k, kt}:
$S=\beta_{Hagedorn} M +\dots$
where $M$ is the black hole mass $M \sim {1\over g^2_s}$.
We have seen that our calculations show that the 0A 2-d black holes
have an entropy $S=4\pi D_h$. Since the string coupling is related to the
dilaton charge at the horizon by $D_h=1/g_s^2$, we find that the black hole
entropy is precisely of the order $1/g_s^2$. It is also interesting to
note that the free energy of type 0A 2-d black hole is non-vanishing
as opposed to the 2-d bosonic string black hole $(\L=0)$. This
suggests that the KKK matrix model might be better suited for
describing the 0A 2d black hole.

\section{Conclusion}\label{conclusion}

We have computed the thermodynamics of 2-d type 0A black holes with
equal number of magnetic and electric D0 branes. Quantities like the ADM
mass and the temperature of the black hole can be computed based on the
geometry of the solution. In this paper we have used a thermodynamical
approach, based on evaluating the thermal partition function, which provides 
new  information
about the solution. We have computed, for the general nonextremal black
hole, its  entropy,  free
energy and chemical potential. An interesting observation is that the
positivity of the ADM mass implies an upper bound on the D0 brane
flux: $q^2\le 16\pi \, e$. Our main results are summarized in table
(\ref{table}). In section \ref{matrix} we have compared our results with
some matrix models that are believed to be of relevance for the type of
2-d black holes we discussed. We found that, as opposed to other
results quoted in the literature,  our results compare
unfavorably with the deformed matrix model proposed by Jevicki and
Yoneya. 
In particular, the form of the thermodynamical entropy of the 
extremal 2-d black hole disagrees with
the matrix model result. On the other hand, we find 
qualitative agreement with the KKK matrix model. In particular, both
entropies go as $g_s^{-2}$. 

The leading term for the entropy of the nonextremal black hole with a
large number of D0 branes goes as $q^2$. We believe this results
captures precisely that the degrees of freedom being described are those
of $q$ D0 branes. This result is very similar to its $AdS_5/CFT_4$
counterpart where the entropy is proportional to $N^2$ describing a
stack of $N$ D3 branes. 

It would be interesting to identify with certainty the matrix model dual
to the black hole solutions discussed here since it will provide a
microscopic basis for our discussion of the thermodynamics.

\section*{Acknowledgments}

It is a pleasure to acknowledge useful discussions with D. Grumiller, A. Jevicki,
I. Klebanov, J. McGreevy, R. McNees, D. Oros  
and  H. Verlinde. LAPZ is partially supported
by DoE grant DE-FG02-95ER40899. D.V. is supported by DOE grant 
DE-FG02-91ER40671.

\appendix

\section{Onshell action}\label{onshellaction}

As it is usual in gravitational Lagrangians, a term with support only on 
the boundary is needed for a well-defined variational problem. Only the 
Einstein-Hilbert term needs a compensating boundary term in the action, 
since it contains second derivatives. The variation of this term
contains a term with second derivatives of the metric variation. This is   
\bea 
\delta I_{EH}^{(2)} & = & \int_{\cal M} \sqrt{-g} e^{-2 \phi}
g^{ab} \delta R_{ab} \nonumber \\                & = & \int_{\cal M}
\sqrt{-g} e^{-2\phi} \nabla^a \left( \nabla^b \delta g_{ab} - g^{bc}
\nabla_a \delta g_{bc} \right) \nonumber \\ & = & \int_{\cal M}
\sqrt{-g} \Big[ \nabla^a \left[ e^{-2\phi} \left( \nabla^b \delta g_{ab}
- g^{bc} \nabla_a \delta g_{bc} \right) \right] \nonumber \\ &&+ 2
\nabla^a e^{-2\phi} \left( \nabla^b \delta g_{ab} - g^{bc} \nabla_a
\delta g_{bc} \right) \Big] \nonumber \\ &=& \int_{\partial {\cal M}}
\sqrt{-h} e^{-2\phi} n^a \Big[ \left( \nabla^b \delta g_{ab} - g^{bc}
\nabla_a \delta g_{bc} \right) \nonumber \\ && +2 \nabla^b \phi \delta
g_{ab} \Big] + \mathrm{bulk\; terms} \nonumber \\ & = & -2 \int_{\partial
{\cal M}} \sqrt{-h} e^{-2\phi} \delta K \label{bndyvari}. 
\eea

\noindent In the last line the second term vanishes 
since $\delta g_{ab} = 0$ on the boundary. In addition the leftover
terms can be shown to give $\delta K$ where $K = h^{ab} \nabla _a n_b$,
where $h_{ab}$ is the metric on the boundary. The bulk terms are simply
dropped.

In order that the overall variation of the action vanishes, we add an
additional boundary term to (\ref{action}) called the Gibbons-Hawking term:

\be
I_{bndy} = 2 \int_{\partial {\cal M}} \sqrt{-h} e^{-2\phi} K \label{bndyaction}
\ee
To calculate thermodynamic quantities, we need to evaluate the thermal 
partition function, which amounts to evaluating the
on-shell action. We can write the on-shell action as a volume
integral piece plus a boundary term. To do this, use the equation of
motion derived by varying with respect to the dilaton:

\be
R + 4 (\nabla \phi)^2 + c + 4 e^{2 \phi} \nabla^a \left( e^{-2\phi} \nabla_a \phi \right) = 0,
\ee
then substitute the above equation of motion into (\ref{action}) to obtain
\bea
I_{bulk}& = &\int_{{\cal M}} \sqrt{-g} 
\left[  - 4 \nabla^a \left( e^{-2\phi} \nabla_a \phi \right) + \Lambda
\right] \nonumber \\
& = & \int_{\cal M} \sqrt{-g} \Lambda - 4 \int_{\partial {\cal M}} 
\sqrt{-h} e^{-2\phi} n^a \nabla_a \phi. 
\eea

Thus the entire on-shell action is 

\be
I_{onshell} = \int_{\cal M} \sqrt{-g} \Lambda 
+2  \int_{\partial {\cal M}} \sqrt{-h} e^{-2\phi} 
\left( K -2 n^a \nabla_a \phi \right). \\
\ee

\section{Marginal tachyon deformations}\label{marginal}

The need to allow for nontrivial tachyon field, that is, $\mu\ne 0$ in
the description of black holes was pointed out by Witten back in
\cite{ew}. The main reason being that many matrix model results are
singular at $\mu=0$. It is, therefore, important to understand the
implications of being forced to work at $\mu=0$ in the context of the
gravity solutions. In this appendix we consider turning on a small
tachyon field in order to shed some light on the structure of the
$\mu=0$ region. 
We thus consider the tachyon which couples to lowest order the other
fields as \cite{0ad}
\be
\int d^2x\sqrt{g}\left(e^{-2\Phi}\left(-(\nabla T)^2
    +\frac{2}{\a'}T^2\right)+ 2\Lambda T^2\right).
\ee
The equation of motion following from the above action is 
\be
\frac{1}{\sqrt{g}}\partial_\a\left(\sqrt{g}e^{-2\Phi}\,
    g^{\a\b}\partial_\b T\right)+ 4(\frac{1}{\a'}e^{-2\Phi}+\L)\, T =0.
\ee
We are interested in solutions of the form $T=T(\phi)$, only depending
on the spatial variable. These solutions determine the profile of
possible perturbations. We consider the $\p\to-\infty$ limit,  that is, we
would like to find out what happens at infinity to small perturbations of
the tachyon field.  In this limit the equation becomes
\be
T''-\sqrt{c}T'+\frac{4}{\a'}T=0,
\ee
and therefore
\be
T\sim e^{\lambda_{\pm} \p}, \quad {\rm where}  \quad
\l_{\pm}=\frac{\sqrt{c}}{2}\left(1\pm \sqrt{1-\frac{16}{c\,\a'}}\right).
\ee
The remarkable result is that both solutions have positive real part and
therefore the tachyon decays at infinity as $\phi \to -\infty$. This
behavior, just as in a similar context discussed in \cite{ew}, signals the existence of a marginal
deformation. This analysis gives us confidence that the solution
described in the paper will exist and perhaps retain some of the
thermodynamical  properties in the presence of a nonzero tachyon
field. This raises hope for the existence of a well-defined $(\mu\ne
0$) matrix
model.

\end{document}